\documentstyle[12pt]{article}
\begin{document}
\newcommand{\mt}{m_{\tau}}
\newcommand{\mpi}{m_{\pi}}
\newcommand{\gva}{\gamma_{VA}}
\newcommand{\spsi}{\sin\psi}
\newcommand{\cpsi}{\cos\psi}
\newcommand{\spsiz}{\sin^{2}\psi}
\newcommand{\cpsiz}{\cos^{2}\psi}
\newcommand{\szpsi}{\sin 2\psi}
\newcommand{\czpsi}{\cos 2\psi}
\newcommand{\sth}{\sin\beta}
\newcommand{\sthz}{\sin^{2}\beta}
\newcommand{\szth}{\sin 2\beta}
\newcommand{\czth}{\cos 2\beta}
\newcommand{\cth}{\cos\beta}
\newcommand{\cthz}{\cos^{2}\beta}
\newcommand{\schi}{\sin\gamma}
\newcommand{\cchi}{\cos\gamma}
\newcommand{\schiz}{\sin^{2}\gamma}
\newcommand{\cchiz}{\cos^{2}\gamma}
\newcommand{\szchi}{\sin 2\gamma}
\newcommand{\czchi}{\cos 2\gamma}
\newcommand{\sph}{\sin\alpha}
\newcommand{\cph}{\cos\alpha}
\newcommand{\ke}{\boldm{K_{1}}}
\newcommand{\kz}{\boldm{K_{2}}}
\newcommand{\kd}{\boldm{K_{3}}}
\newcommand{\kv}{\boldm{K_{4}}}
\newcommand{\kf}{\boldm{K_{5}}}
\newcommand{\ks}{\boldm{K_{6}}}
\newcommand{\keb}{\boldm{\overline{K}_{1}}}
\newcommand{\kzb}{\boldm{\overline{K}_{2}}}
\newcommand{\kdb}{\boldm{\overline{K}_{3}}}
\renewcommand{\baselinestretch}{1.}
 
\begin{center}
{{\bf STRUCTURE FUNCTIONS AND DISTRIBUTIONS IN\\
 SEMILEPTONIC TAU DECAYS}}\\[2mm]

{J.~H.~K\"UHN\\
{\em Institut f\"ur Theoretische Teilchenphysik,
Universit\"at Karlsruhe, Kaiserstr.12,\\ 76128 Karlsruhe, Germany. }\\
\vspace{0.3cm}
and\\
\vspace*{0.3cm}
E. MIRKES\footnote{Talk
  presented by E. Mirkes at the DPF94 Meeting, Albuquerque,
New Mexico, USA; August 2-6, 1994.}\\
{\em Physics Department, University of Wisconsin, Madison,\\
WI 53706, USA }}
\end{center}
\setlength{\baselineskip}{2.6ex}
 
\begin{center}
\parbox{13.0cm}
{\begin{center} ABSTRACT \end{center}
{\small \hspace*{0.3cm}
Semileptonic decays of polarized $\tau$ leptons
are investigated. The most general angular distribution
of three meson final states
 ($\tau\rightarrow \pi\pi\nu,\,K\pi\nu,\,
 \pi\pi\pi\nu,\\ \,K\pi\pi\nu,\,
KK\pi\nu,\, KKK\nu,\, \eta\pi\pi\nu,\,\ldots{}$)
is discussed. It is shown, that the most general 
distribution can be characterized by 16 structure functions, most of which
can be determined in currently ongoing high statistics experiments.
Emphasis is put on $\tau$ decays in $e^{+}e^{-}$ experiments
where the neutrino escapes detection and the $\tau$ rest frame
cannot be reconstructed.
The structure of the hadronic matrix elements, based on CVC and
chiral lagrangians, is discussed.
}}
\end{center}
\vspace{-13.cm}
\hbox to \hsize{
\hskip.5in \raise.1in\hbox{\bf University of Wisconsin - Madison}
\hfill$\vcenter{\hbox{\bf MAD/PH/848}
            }$
               }
\mbox{}
\hfill  August  1994   \\[12.9cm]
With the experimental progress in $\tau$-decays an ideal tool for
studying strong interaction physics has been developed. 
In this paper we show, that  detailed informations about the hadronic 
charged current for the decay into three pseudoscalar mesons
can be derived from the study of angular distributions.
Consider the semileptonic $\tau$-decay
\vspace{-2mm}
  \begin{equation}
\tau(l,s)\rightarrow\nu(l^{\prime},s^{\prime})
+h_{1}(q_{1},m_{1})+h_{2}(q_{2},m_{2})+h_{3}(q_{3},m_{3})\>, \label{process}
  \end{equation}
\vspace{-2mm}
where $h_i(q_i,m_i)$ are pseudoscalar mesons.
The  matrix element reads as 
  \begin{equation}
{\cal{M}}=\frac{G}{\sqrt{2}}\,\bigl(^{\cos\theta_{c}}_{\sin\theta_{c}}\bigr)
\,M_{\mu}J^{\mu}\>,
\label{mdef}
  \end{equation}
with $G$ the Fermi-coupling constant. The  cosine and the sine of the 
Cabibbo angle ($\theta_C$) 
in (\ref{mdef}) have to be used for Cabibbo allowed $\Delta S=0$ and 
Cabibbo suppressed $|\Delta S|=1$ decays, respectively.
The leptonic ($M_\mu$) and hadronic ($J^\mu$) currents are given by 
$
M_{\mu}=
\bar{u}(l^{\prime},s^{\prime})\gamma_{\mu}(g_{V}-g_{A}\gamma_{5})u(l,s)      
$
and
$
J^{\mu}(q_{1},q_{2},q_{3})=\langle h_{1}(q_{1})h_{2}(q_{2})h_{3}(q_{3})
|V^{\mu}(0)-A^{\mu}(0)|0\rangle.
$
$V^\mu$ and $A^\mu$ are the vector and axial vector quark currents, 
respectively. 
The most general ansatz for the matrix element of the
quark current $J^{\mu}$  
is characterized by four formfactors \cite{km1}
  \begin{eqnarray}
J^{\mu}(q_{1},q_{2},q_{3})
&=&   V_{1}^{\mu}\,F_{1}
    + V_{2}^{\mu}\,F_{2}
    +\,i\, V_{3}^{\mu}\,F_{3}   
    + V_{4}^{\mu}\,F_{4}\label{f1234}\>,
  \end{eqnarray}
with
  \begin{equation}
    \begin{array}{ll}
V_{1}^{\mu}&=q_{1}^{\mu}-q_{3}^{\mu}-Q^{\mu}\frac{Q(q_{1}-q_{3})}{Q^{2}}\>,
\\[2mm]
V_{2}^{\mu}&=q_{2}^{\mu}-q_{3}^{\mu}-Q^{\mu}\frac{Q(q_{2}-q_{3})}{Q^{2}}\>,
\\[2mm]
V_{3}^{\mu}&= \epsilon^{\mu\alpha\beta\gamma}q_{1\,\alpha}q_{2\,\beta}
                                            q_{3\,\gamma}\>,
\\[2mm]
V_{4}^{\mu}&=q_{1}^{\mu}+q_{2}^{\mu}+q_{3}^{\mu}\,=Q^{\mu}\>.
    \end{array}
  \end{equation} 
The formfactors $F_{1}$ and $F_{2}(F_{3})$ originate from the axial vector
hadronic current (vector current) and correspond to spin 1,
whereas $F_{4}$  is due to the spin zero part of the axial current matrix
element.
The formfactors $F_1$  and $F_2$ can be predicted by chiral lagrangians,
supplemented by informations about resonance parameters.
Parametrizations for the $3\pi$ final states
can be found in refs. [1-3].
In this case, only the axial vector current formfactors $F_1$ and $F_2$
contribute due to the $G$ parity of the pions.
The $3\pi$ decay mode offers a unique tool  for the study of 
$\rho,\rho'$ resonance parameters competing well with low
energy $e^+e^-$ colliders with energies in the region below 1.7 GeV.
As we will see later, the two body ($\rho$ and $\rho'$) resonances 
can be fixed by taking
ratios of hadronic structure functions, whereas 
the measurement of four structure functions can be used to
put constraints on the $a_1$ parameters.
The decay modes involving
different mesons (for example $K\pi\pi,\,KK\pi$ or $\eta\pi\pi$)
allow for axial and vector current contributions at the same time.
Explicit parametrizations for the form factors in
 these decay modes are presented in refs. [4,5].
The vector formfactor $F_3$ is related to the Wess-Zumino anomaly \cite{wz},
whereas the axial vector form factors are again predicted
by chiral lagrangians. The latter decay modes allow also
for the study of $J^{PC}=0^{-+}$ and
$J^{PC}=1^{++}$ resonances which are not directly accessible from 
other experiments.

Let us now introduce the formalism of the hadronic structure functions.
The differential decay rate  is obtained from
  \begin{equation}
d\Gamma(\tau\rightarrow \nu_\tau\,3h)=\frac{1}{2\mt} 
\frac{G^{2}}{2}\,\bigl(^{\cos^2\theta_{c}}_{\sin^2\theta_{c}}\bigr)\,
\left\{L_{\mu\nu}H^{\mu\nu}\right\}
\,d\mbox{PS}^{(4)}\>,
 \label{decay}
  \end{equation}
where $L_{\mu\nu}=M_\mu (M_\nu)^\dagger$ and 
$H^{\mu\nu}\equiv J^{\mu}(J^{\nu})^{\dagger}$.
The considered decays (\ref{process}) are most easily analyzed in the 
hadronic rest frame
$\vec{q}_{1}+\vec{q}_{2}+\vec{q}_{3}=\vec{Q}=0$.
The orientation of the hadronic
system is characterized by 
three Euler angles ($\alpha,\beta$ and $\gamma$) as introduced in
refs. [1,3].
Performing the analysis of $\tau \to \nu_\tau $+ 3 mesons in the 
hadronic rest frame  has the advantage that the product of the
hadronic  
and the leptonic tensors reduce to a sum \cite{km1}
\begin{equation} L^{\mu\nu}H_{\mu\nu}=\sum_{X} L_XW_X.
\label{eq1}
\end{equation}
In fact in this system the hadronic tensor $H^{\mu\nu}$ is decomposed into 16
hadronic structure
functions $W_{X}$
corresponding to 16 density matrix elements for a hadronic system in a spin one
[$V_1^{\mu}, V_2^{\mu}, V_3^{\mu}$]
and spin zero state $[V_4^{\mu}]$ 
(nine of them originate from a pure spin one and the
remaining are pure spin zero  or interference terms).
 The 16 structure functions
 contain the
dynamics of the three meson decay and 
  depend only   on the hadronic invariants $Q^2$ and
the Dalitz plot variables $s_{i}$.
The leptonic factors
 $L_X$ contain the dependence on the Euler angles,
( which 
determine the orientation of the hadronic system),
on the $\tau$ polarization, 
on the chirality parameter $\gamma_{VA}$
 and on the total energy of the hadrons in
the laboratory frame as well.
Analytical expressions for the 16 coefficients $L_X$ are first presented
in ref. [1].
They can also be applied to the case,
 where the $\tau$
 rest frame cannot be reconstructed
because of the unknown neutrino momentum. 
The dependence  of these  coefficients on the  $\tau$ polarization 
allow for an improved measurement of the $\tau$ polarization at LEP
[see for example refs. [7,8] and references therein].

The hadronic structure functions $W_{X}$
on the other hand contain the full dynamics of the hadronic decay and a 
measurement of these structure functions provide a unique tool for low enery
hadronic physics.
They can  be calculated from a decomposition of the hadronic matrix
element $J^{\mu}$ and can be expressed in terms 
of the form factors $F_i$.
We list here only the result for the pure spin one state. 
  \begin{eqnarray}  \hspace{3mm}
W_{A}  &=&   \hspace{3mm}(x_{1}^{2}+x_{3}^{2})\,|F_{1}|^{2}
           +(x_{2}^{2}+x_{3}^{2})\,|F_{2}|^{2}
           +2(x_{1}x_{2}-x_{3}^{2})\,\mbox{Re}\left(F_{1}F^{\ast}_{2}\right)
                             \>,      \nonumber \\[3mm]
W_{B}  &=& \hspace{3mm} x_{4}^{2}|F_{3}|^{2}
                                \>,   \nonumber \\[3mm]
W_{C}  &=&  \hspace{3mm} (x_{1}^{2}-x_{3}^{2})\,|F_{1}|^{2}
           +(x_{2}^{2}-x_{3}^{2})\,|F_{2}|^{2}
           +2(x_{1}x_{2}+x_{3}^{2})\,\mbox{Re}\left(F_{1}F^{\ast}_{2}\right)
                                \>,   \nonumber \\[3mm]
W_{D}  &=&  \hspace{3mm}2\left[ x_{1}x_{3}\,|F_{1}|^{2}
           -x_{2}x_{3}\,|F_{2}|^{2}
           +x_{3}(x_{2}-x_{1})\,\mbox{Re}\left(F_{1}F^{\ast}_{2}\right)\right]
                               \>,    \nonumber \\[3mm]
W_{E}  &=& -2x_{3}(x_{1}+x_{2})\,\mbox{Im}\left(F_{1}
                    F^{\ast}_{2} \right) \label{walldef}\>,\\[3mm]
W_{F}  &=&  \hspace{3mm}
          2x_{4}\left[x_{1}\,\mbox{Im}\left(F_{1}F^{\ast}_{3}\right)
                     + x_{2}\,\mbox{Im}\left(F_{2}F^{\ast}_{3}\right)\right]
                               \>,    \nonumber \\[3mm]
W_{G}  &=&- 2x_{4}\left[x_{1}\,\mbox{Re}\left(F_{1}F^{\ast}_{3}\right)
                     + x_{2}\,\mbox{Re}\left(F_{2}F^{\ast}_{3}\right)\right]]
                            \>,       \nonumber \\[3mm]
W_{H}  &=& \hspace{3mm}
      2x_{3}x_{4}\left[\,\mbox{Im}\left(F_{1}F^{\ast}_{3}\right)
                     -\,\mbox{Im}\left(F_{2}F^{\ast}_{3}\right)\right]
                              \>,     \nonumber \\[3mm]
W_{I}  &=&- 2x_{3}x_{4}\left[\,\mbox{Re}\left(F_{1}F^{\ast}_{3}\right)
                     -\,\mbox{Re}\left(F_{2}F^{\ast}_{3}\right)\right]
                                 .  \nonumber 
\end{eqnarray}
The remaining structure functions originating 
from a possible (small) contribution
from a spin zero state are presented  in [1].
The variables $x_i$ are defined by
$
x_{1}= V_{1}^{x}=q_{1}^{x}-q_{3}^{x},\,
x_{2}= V_{2}^{x}=q_{2}^{x}-q_{3}^{x},\,
x_{3}= V_{1}^{y}=q_{1}^{y}=-q_{2}^{y},\,
x_{4}= V_{3}^{z}=\sqrt{Q^{2}}x_{3}q_{3}^{x},\,
$
where $q_i^{x}$ ($q_i^{y}$) denotes the $x$ ($y$) component of the momentum of 
meson $i$ in the hadronic rest frame.
The structure functions can be extracted by taking suitable moments
with respect to an appropriate product of two Euler angles
\cite{km1}.
An alternative method to extract the structure functions can be achieved
by a direct fit to the expressions (\ref{walldef}) \cite{jayant}.

As an example, we will now present numerical results for the 
non vanishing structure functions $W_A, W_C, W_D$ and $W_E$
in the $3\pi$ decay mode, which originates from the spin one part 
of the hadronic current.
\begin{figure}[tbh]
\vspace{7cm}
\caption{Ratio of the spin one hadronic structure functions
$w_C/w_A,w_D/w_A,w_E/w_A$ (from top to bottom) for 
$\tau\rightarrow \nu\pi\pi\pi $ as a function of $Q^2$.}
\end{figure}
Figure 1 shows predictions for the structure function ratios
$w_C/w_A, w_D/w_A$ and $w_E/w_A$ as a function of $Q^2$, where
we have integrated over the Dalitz plot variables $s_i$ [The
integrated structure functions are denoted by lower case letter $w_X$].
The results are based on the same parametrization of
the formfactors as used in [1].
Although we have lost informations on the resonance
parameters in the  two body decays by integrating over $s_1$ and $s_2$,
we observe interesting structures\footnote{More constraints on the
two body resonances can be obtained by analyzing the full 
dependence on $Q^2$ and $s_i$, which should be accessible with the 
present  high statistic experiments.}.
One observes that all normalized structure functions are  sizable.
$w_{C}/w_{A}$ approaches its maximal value $1$ for small $Q^{2}$.
Note, that the dependence on the $a_1$ mass and width parameters
cancel in the ratio $w_X/w_A$ in fig.~1.
\begin{figure}[tbh] 
\vspace{7cm}
\caption{Spin one hadronic structure functions
$w_A,w_C,w_D,w_E$ (from top to bottom) for
$\tau\rightarrow \nu\pi\pi\pi$
as a function of $Q^2$.
Results are shown for two sets of $a_1$ parameters:
$m_{a_1}=1.251$ GeV, $\Gamma_{a_1}= 0.599$ GeV (solid)
and
$m_{a_1}=1.251$ GeV, $\Gamma_{a_1}= 0.550$ GeV (dashed) }
\end{figure}
On the other hand, the $Q^2$ distributions of the structure functions
$w_{A,C,D,E}$ presented in fig.~2 are very sensitive to the $a_1$ parameters.
As an example, fig. 2 shows predictions for the structure functions,
where two different values for the $a_1$ width has been used.
Therefore, the ratios in fig.~1 can be used to fix the model dependence
in the two body resonances, whereas the structure functions itself
put then rigid constraints on the $a_1$ parameters. It is therefore possible
to test the hadronic physics in much more detail than it is possible 
by rate measurements alone.

The technique of structure functions also allow for a model independent
test of the presence of spin zero components in the hadronic current.
Such a contribution would lead to  additional structure functions to 
(\ref{walldef}) originating from the interference with the (large)
spin one contributions\cite{km1}.

A detailed discussion of the matrix elements for
the decay modes involving different pseudo scalar mesons 
[$K\pi\pi\nu,\, KKK\nu,\, \eta\pi\pi\nu$] 
 together with predictions
for the corresponding structure functions and angular distributions
is presented in ref. [4].
In this case, all 9 structure functions in (\ref{walldef}) are
nonvanishing because of the interference of the anomaly with
the axial vector contributions.
An analyses of these distributions would  allow to test
the underlying  hadronic physics [like the different contributions from 
the axial and vector (Wess-Zumino anomaly) current] in detail.
It is thus possible to confirm  (not only qualitatively)
the presence of the Wess-Zumino anomaly in the decays modes
 $\tau\rightarrow \nu K\pi\pi$ and
$\tau\rightarrow \nu K\pi\pi$.

In ref. [9], the technique of the structure functions
has been  extended to the $\tau\rightarrow\nu\omega\pi$ decay mode.
This allows to test
the model for the hadronic matrix element in this decay mode,  which involves 
both a vector and a second class axial vector current.

\bibliographystyle{unsrt}

\end{document}